\documentstyle[aps]{revtex} 
\title{Can Science `explain' Consciousness ?} 
\author{\sf M. K. Samal} 
\address{ Non-Accelerator Particle Physics (NAPP) Group,\\ Indian
Institute of Astrophysics, Bangalore-560 034, India.}
\date{e-mail: {\sf mks@iiap.ernet.in}} 
\begin{document} 
\maketitle 
\begin{abstract} 
Consciousness is the process by which one attributes `meaning' to the
world. Considering F$\phi$llesdal's definition of `meaning' as the joint
product of all `evidence' that is available to those who `communicate', we
conclude that science can, not only reduce all the {\em evidence} to a
Basic Entity (we call BE), but also can `explain' consciousness once a
suitable definition for {\em communication} is found that exploits the
quantum superposition principle to incorporate the fuzzyness of our
experience. Consciousness may be beyond `computability', but it is not
beyond `communicability'. 
\end{abstract} 
\sf
\section{Introduction}
Among all the human endeavours, science can be considered to be the most
powerful for the maximum power it endowes us to manipulate the nature
through an understanding of our position in it. This understanding is
gained when a set of careful observations based on tangible perceptions,
acquired by sensory organs and/or their extensions, is submitted to the
logical analysis of human intellect as well as to the intuitive power of
imagination to yield the abstract fundamental laws of nature that are not
self-evident at the gross level of phenomenal existence. There exists a
unity in nature at the level of laws that corresponds to the manifest
diversity at the level of phenomena. 

Can consciousness be understood in this sense by an appropriate use of the
methodology of science ? The most difficult problem related to
consciousness is perhaps, `how to define it ?'. Consciousness has remained
a unitary subjective experience, its various `components' being reflective
(the recognition by the thinking subject of its own actions and mental
states), perceptual (the state or faculty of being mentally aware of
external environment) and a free will (volition). But how these components
are integrated to provide the unique experience called `consciousness',
familiar to all of us, remains a mystery. Does it lie at the level of
`perceptions' or at the level of `laws' ? {\it Can it be reduced to some
basic `substance' or `phenomenon' ? Can it be manipulated in a controlled
way ? Is there a need for a change of either the methodology or the
paradigm of science to answer the above questions ?} In this article, I
make a modest attempt to answer these questions, albeit in a speculative
manner. 
\section{Can Consciousness be reduced further ?}
Most of the successes of science over the past five hundred years or so 
can be attributed to the great emphasis it lays on the `reductionist 
paradigm'. Following this approach, can consciousness be reduced either to 
`substance' or `phenomena' in the sense that by understanding which one 
can understand consciousness ?
\subsection{Physical Substratum}
The attempts to reduce consciousness to a physical basis have been made 
in the following ways by trying to understand the mechanism 
and functioning of the human brain in various different contexts.
\begin{itemize}
\item{\bf Physics}

The basic substratum of physical reality is the `state' of the system and
the whole job of physics can be put into a single question : `given the
initial state, how to predict its evolution at a later time ?'. In
classical world, the state and its evolution can be reduced to events and
their spatio-temporal correlations. Consciousness has no {\it direct} role
to play in this process of reduction, although it is responsible to find an
`objective meaning' in such a reduction. 

But the situation is quite different in the quantum world as all relevant
physical information about a system is contained in its wavefunction (or
equivalently in its state vector), which is not physical in the sense of
being directly measurable.  Consciousness plays no role in the
deterministic and unitary Schr\"odinger evolution (i.e. the {\bf
U}-process of Penrose\cite{penrose}) that the `unphysical' wavefunction
undergoes. 

To extract any physical information from the wavefuction one has to use
the Born-Dirac rule and thus probability enters in a new way into the
quantum mechanical description despite the strictly deterministic nature
of evolution of the wavefunction. The measurement process forces the
system to choose an `actuality' from all `possibilities' and thus leads to
a non-unitary collapse of the general wavefunction to an eigenstate
(i.e.  the {\bf R}-process of Penrose\cite{penrose}) of the concerned
observable.  The dynamics of this {\bf R}-process is not known and it is
here some authors like Wigner have brought in the consciousness of the
observer to cause the collapse of the wavefunction.  But instead of
explaining the consciousness, this approach uses consciousness for the
sake of Quantum Mechanics which needs the {\bf R}-process along with the
{\bf U}-process to yield all its spectacular successes. 

The {\bf R}-process is necessarily {\it non-local} and is governed by an
irreducible element of chance, which means that the theory is not
naturalistic: the dynamics is controlled in part by something that is not
a part of the physical universe. Stapp\cite{stapp} has given a quantum
mechanical model of the brain dynamics in which this quantum selection
process is a causal process governed not by pure chance but rather by a
mathematically specified non-local physical process identifiable as the
{\it conscious process}.  It was reported\cite{ranjit} that attempts have
been made to explain consciousness by relating it to the `quantum events',
but any such attempt is bound to be futile as the concept of `quantum
event' in itself is ill-defined ! 

Keeping in view the fundamental role that the quantum vacuum plays in
formulating the quantum field theories of all four known basic
interactions of nature spreading over a period from the big-bang to the
present, it has been suggested\cite{sreek} that if at all consciousness be
reduced to anything `fundamental' that should be the `quantum vacuum' in
itself. But in such an approach the following questions arise: 1) If
consciousness has its origin in the quantum vacuum that gives rise to all
fundamental particles as well as the force fields, then why is it that
only living things possess consciousness ?, 2) What is the relation
between the quantum vacuum that gives rise to consciousness and the
space-time continuum that confines all our perceptions through which
consciousness manifests itself ?, 3) Should one attribute consciousness
only to systems consisting of `real' particles or also to systems containing
`virtual' particles ?  Despite these questions, the idea of tracing
the origin of `consciousness' to `substantial nothingness' appears quite
promising because the properties of `quantum vacuum' may ultimately
lead us to an understanding of the dynamics of the {\bf R}-process and thus 
to a physical comprehension of consciousness.

One of the properties that distinguishes living systems from the
non-living systems is their ability of self-organisation and complexity. 
Since life is a necessary condition for possessing consciousness, can one
attribute consciousness to a `degree of complexity' in the sense that
various degrees of consciousness can be caused by different levels of
complexity? Can one give a suitable quantitative definition of
consciousness in terms of `entropy' that describes the `degree of
self-organisation or complexity' of a system ?  What is the role of
non-linearity and non-equilibrium thermodynamics in such a definition of
consciousness ?  In this holistic view of consciousness what is the
role played by the phenomenon of {\it quantum non-locality}, first
envisaged in EPR paper\cite{epr} and subsequently confirmed
experimentally\cite{aspect} by Aspect et. al ? What is the role of
irreversibility and dissipation in this holistic view ? 
\item{\bf Neuro-biology}

On the basis of the vast amount of information available on the structure
and the modes of communication (neuro-transmitters, neuro-modulators,
neuro-hormones) of the neuron, neuroscience has empirically
found\cite{sobhini} the neural basis of several attributes of
consciousness.  With the help of modern scanning techniques and by direct
manipulations of the brain, neuro-biologists have found out that various
human activities (both physical and mental) and perceptions can be mapped
into almost unique regions of the brain. Awareness, being intrinsic to
neural activity, arises in higher level processing centers and requires
integration of activity over time at the neuronal level.  But there exists
no particular region that can be attributed to have given rise to
consciousness. Consciousness appears to be a collective phenomena where
the `whole' is much more than the sum of parts ! {\em Is each neuron having 
the `whole of consciousness' within it, although it does work towards a 
particular attribute of consciousness at a time ?}

Can this paradigm of finding neural correlates of the attributes of
consciousness be fruitful in demystifying consciousness ? Certainly not ! 
As it was aptly concluded\cite{tondon} the currently prevalent
reductionist approaches are unlikely to reveal the basis of such holistic
phenomenon as consciousness. There have been holistic
attempts\cite{hameroff,penrose} to understand consciousness in terms of
collective quantum effects arising in cytoskeletons and microtubles;
minute substructures lying deep within the brain's neurons.  The effect of
{\it general anaesthetics} like chloroform (CHCl$_3$), isofluorane
(CHF$_2$OCHClCF$_3$) etc. in swiching off the consciousness, not only in
higher animals such as mammals or birds but also in paramecium, amoeba,
or even green slime mould has been advocated\cite{hamerwatt} to be
providing a direct evidence that the phenomenon of consciousness is
related to the action of the cytoskeleton and to microtubles. But all
the implications of `quantum coherence' regarding consciousness in such
approach can only be unfolded after we achieve a better understanding of
`quantum reality', which still lies ahead of the present-day physics. 
\item{\bf Artificial Intelligence}

Can machines be intelligent ? Within the restricted definition of
`artificial intelligence', the neural network approach has been the most
promising one. But the possibility of realising a machine capable of
artificial intelligence based on this approach is constrained at
present\cite{vidya} by the limitations of `silicon technology' for
integrating the desired astronomical number of `neuron-equivalents' into a
reasonable compact space. Even though we might achieve such a feat in the
foreseeable future by using chemical memories, it is not quite clear
whether such artificially intelligent machines can be capable of
`artificial consciousness'. Because one lacks at present a suitable
working definition of `consciousness' within the frame-work of studies
involving artificial intelligence. 

Invoking G\"odel's incompleteness theorem, Penrose has
argued\cite{penrose} that the technology of electronic computer-controlled
robots will not provide a way to the artificial construction of an
actually intelligent machine--in the sense of a machine that `understands'
what it is doing and can act upon that understanding.  He maintains that
human understanding (hence consciousness) lies beyond formal arguments
and beyond computability i.e. in the Turing-machine-accessible sense. 

Assuming the inherent ability of quantum mechanics to incorporate
consciousness, can one expect any improvement in the above situation by
considering `computation' to be a physical process that is governed by the
rules of quantum mechanics rather than that of classical physics ?  In
`Quantum computation'\cite{qcomp} the classical notion of a Turing machine
is extended to a corresponding quantum one that takes into account {\it
the quantum superposition principle}. In `standard' quantum computation,
the usual rules of quantum theory are adopted, in which the system evolves
according to the {\bf U}-process for essentially the entire operation, but
the {\bf R}-process becomes relevant mainly only at the end of the
operation, when the system is `measured' in order to ascertain either the
termination or the result of the computation. 

Although the superiority of the quantum computation over classical
computation in the sense of complexity theory have been
shown\cite{qcmfast}, Penrose insists that it is still a `computational'
process since {\bf U}-process is a computable operation and {\bf
R}-process is purely probabilistic procedure. What can be achieved in
principle by a quantum computer could also be achieved, in principle, by a
suitable Turing-machine-with-randomiser. Thus he concludes that even a
quantum computer would not be able to perform the operations required for
human conscious understanding. But we think that such a view is limited
because `computation' as a process need not be confined to a
Turing-machine-accessible sense and in such situations one has to explore
the power of quantum computation in understanding consciousness.
\end{itemize}

We conclude from the above discussions that the basic physical substrata
to which consciousness may be reduced are `neuron', `event' and `bit' at
the classical level, whereas at the quantum level they are `microtuble',
`wavefunction' and `qubit'; depending on whether the studies are done in
neuro-biology, physics and computer science respectively.  Can there be a
common platform for these trio of substrata ? 

We believe the answer to be in affirmative and the first hint regarding
this comes from Wheeler's\cite{wheeler} remarkable idea: `` {\bf it from
bit} {\it i.e.} every {\bf it} -- every particle, every field of force,
even the spacetime continuum itself -- derives its function, its meaning,
its very existence entirely -- even if in some contexts indirectly -- from
the apparatus-elicited answers to yes or no questions, binary choices,
{\bf bits}".  This view of the world refers not to an object, but to a
vision of a world derived from pure logic and mathematics in the sense
that an {\em immaterial} source and explanation lies at the bottom of
every item of the physical world. In a recent report\cite{wilczek} the
remarkable extent of embodiment of this vision in modern physics has beed
discussed alongwith the possible difficulties faced by such a scheme.  But
can this scheme explain consciousness by reducing it to {\bf bits} ?
Perhaps not unless it undergoes some modification. Why ? 

Because consciousness involves an awareness of an endless mosaic of 
qualitatively different things --such as the colour of a rose, the 
fragrance of a perfume, the music of a piano, the tactile sense of 
objects, the power of abstraction, the intuitive feeling for time and 
space, emotional states like love and hate, the ability to put oneself in 
other's position, the abilitiy to wonder, the power to wonder at one's 
wondering etc. It is almost impossible to reduce them all to the 0-or-1 
sharpness of the definition of `bits'. A major part of human experience 
and consciousness is {\it fuzzy} and hence can not be reduced to yes or 
no type situations. Hence we believe that `bit' has to be modified to 
incorporate this fuzzyness of the world. Perhaps the quantum 
superposition inherent to a `qubit' can help. Can one then reduce the 
consciousness to a consistent theory of `quantum information' based on 
qubits ? Quite unlikely, till our knowledge of `quantum reality' and 
the `emergence of classicality from it' becomes more clear.

The major hurdles to be cleared are (1) {\bf Observer or Participator ?}
(In such equipment-evoked, quantum-information-theoretic approach, the
inseparability of the observer from the observed will bring in the quantum
measurement problem either in the form of dynamics of the {\bf R}-process
or in the emergence of classicality of the world from a quantum
substratum. We first need the solutions to these long-standing problems
before attempting to reduce the `fuzzy' world of consciousness to
`qubits'! ); (2) {\bf Communication ?} (Even if we get the solutions to
the above problems that enable us to reduce the `attributes of
consciousness' to `qubits', still then the `dynamics of the process that
gives rise to consciousness' will be beyond `quantum information' as it
will require a suitable definition of `communication' in the sense
expressed by F$\phi$llesdal\cite{follesdal} `` Meaning is the joint
product of all evidence that is available to those who communicate''.
Consciousness helps us to find a `meaning' or `understanding' and will
depend upon `communication'. Although all `evidence' can be reduced to
qubits, `communication' as an exchange of qubits has to be well-defined.
Why do we say that a stone or a tree is unconscious ? Is it because we do
not know how to `communicate' with them ? Can one define `communication'
in physical terms beyond any verbal or non-verbal language ? Where does
one look for a suitable definition of `communication' ? Maybe one has to
define `communication' at the `substantial nothingness' level of quantum
vacuum.); (3) {\bf Time's Arrow ?} (How important is the role of memory in
`possessing consciousness' ?  Would our consciousness be altered if the
world we experience were reversible with respect to time ? Can our
consciousness ever find out why it is not possible to influence the past
?). 

Hence we conclude that although consciousness may be beyond
`computability', it is not beyond `quantum communicability' once a
suitable definition for `communication' is found that exploits the quantum
superposition principle to incorporate the fuzzyness of our experience. 
Few questions arise: 1) how to modify the qubit ?, 2) can a suitable
definition of `communication', based on immaterial entity like `qubit' or
`modified qubit', take care of non-physical experience like dream or
thoughts ?  We assume, being optimistic, that a suitable modification of
`qubit' is possible that will surpass the hurdles of communicability,
dynamics of {\bf R}-process and irreversibility. For the lack of a better
word we will henceforth call such a modified qubit as `Basic Entity' (BE). 
 
\subsection{Non-Physical Substratum}

Unlike our sensory perceptions related to physical `substance' and
`phenomena' there exists a plethora of human experiences like dreams,
thoughts and lack of any experience during sleep which are believed to be
non-physical in the sense that they cannot be reduced to anything basic
within the confinement of space-time and causality. For example one cannot
ascribe either spatiality or causality to human thoughts, dreams etc. Does
one need a frame-work that transcends spatio-temporality to incorporate
such non-physical `events' ? Or can one explain them by using BE ? The
following views can be taken depending on one's belief: 
\begin{itemize}
\item{\bf Modified BE [ M(BE) ]}

What could be the basic substratum of these non-physical entities ? Could
they be understood in terms of any suitably modified physical substratum ?
At the classical level one might think of reducing them to `events' which,
unlike the physical events, do not have any reference to spatiality. 
Attempts\cite{nsingh} have been made to understand the non-physical
entities like thoughts and dreams in terms of temporal events and
correlation between them. Although such an approach may yield the
kinematics of these non-physical entities, it is not clear how their
dynamics i.e. evolution etc. can be understood in terms of temporal
component alone without any external spatial input, when in the first
place they have arose from perceptions that are meaningful only in the
context of spatio-temporality ?! Secondly, it is not clear why the `mental
events' constructed after dropping the spatiality should require new set
of laws that are different from the usual physical laws. 

At the quantum level one might try to have a suitable modification of the
wavefunction to incorporate these non-physical entities. One may make the
wavefunction depend on extra parameters\cite{kaushal}, either physical or
non-physical, to give it the extra degrees of freedom to mathematically
include more information. But such a wavefunction bound to have severe
problems at the level of interpretation. For example, if one includes an
extra parameter called `meditation' as a new degree of freedom apart from
the usual ones, then how will one interpret squared modulus of the
wavefunction ? It will be certainly too crude to extend the Born rule to
conclude that the squared modulus in this case will give the probability
of finding a particle having certain meditation value ! Hence this
kind of modification will not be of much help except for the apparent
satisfaction of being able to write an eigenvalue equation for dreams or
emotions ! This approach is certainly not capable of telling how the
wavefunction is related to consciousness, let alone a mathematical
equation for the evolution of consciousness !

If one accepts consciousness as a phenomenon that arises out of execution
of processes then any suggested\cite{rsingh} new physical basis can be
shown to be redundant.  As we have concluded earlier, all such possible
processes and their execution can be reduced to BE and spatio-temporal
correlations among BE using a suitable definition of communication. 

Hence to incorporate non-physical entities as some kind of information one
has to modify the BE in a subtle way. Schematically M(BE)= BE $\otimes$ X,
where $\otimes$ stands for a yet unknown operation and X stands for
fundamental substratum of non-physical information. X has to be different
from BE;  otherwise it could be reducible to BE and then there will be no
spatio-temporal distinction between physical and non-physical information. 
But, how to find out what is X ? Is it evident that the laws for M(BE)
will be different from that for BE ? 
\item{\bf Give up BE}

One could believe that it is the `Qualia' that constitutes consciousness
and hence consciousness has to be understood at a phenomenological level
without disecting it into BE or M(BE). One would note that consciousness
mainly consists of three phenomenological processes that can be roughly
put as retentive, reflective and creative. But keeping the tremendous
progress of our physical sciences and their utility to neuro-sciences in
view, it is not unreasonable to expect that all these three
phenomenological processes, involving both human as well as
animal\cite{sinha} can be understood oneday in terms of M(BE). 
\item{\bf Platonic BE}

It has been suggested\cite{sarukai} that consciousness could be like
mathematics in the sense that although it is needed to comprehend the
physical reality, in itself it is not `real'. 

The `reality' of mathematics is a controversial issue that brings in the
old debate between the realists and the constructivists whether a
mathematical truth is `a discovery' or `an invention' of the human mind ?
{\em Should one consider the physical laws based on mathematical truth as
real or not ?!} The realist's stand of attributing a Platonic existence to
the mathematical truth is a matter of pure faith unless one tries to get
the guidance from the knowledge of the physical world. It is doubtful
whether our knowledge of physical sciences provides support for the
realist's view if one considers the challenge to `realism' in physical
sciences by the quantum world-view, which has been substantiated in recent
past by experiments\cite{aspect} that violate Bell's inequalities. 

{\em Even if one accepts the Platonic world of mathematical forms, this no
way makes consciousness non-existent or unreal. Rather the very fact that
truth of such a platonic world of mathematics yields to the human
understanding as much as that of a physical world makes consciousness all
the more profound in its existence}. 
\end{itemize}
\section{Can Consciousness be manipulated ?}

Can consciousness be manipulated in a controlled manner ?  Experience
tells us how difficult it is to control the thoughts and how improbable it
is to control the dreams.  We discuss below few methods prescribed by
western psycho-analysis and oriental philosophies regarding the
manipulation of consciousness. Is there a lesson for modern science to
learn from these methods ? 
\subsection{Self}

The subject of `self' is usually considered to belong to an `internal space'
in contrast to the external space where we deal with others. We will
consider the following two cases here: 
\begin{itemize}
\item{\bf Auto-suggestions}

There have been evidences that by auto-suggestions one can control one's
feelings like pain and pleasure. Can one cure oneself of diseases of
physical origin by auto-suggestions ? This requires further
investigations. 
\item{\bf Yoga and other oriental methods}

The eight-fold ({\it asthanga}) Yoga of Patanjali is perhaps the most
ancient method prescribed\cite{bks} to control one's thought and to direct
it in a controlled manner. But it requires certain control over body and
emotions before one aspires to gain control over mind. In particular it
lays great stress on `breath control' ({\it pranayama}) as a means to
relax the body and to still the mind. In its later stages it provides
systematic methods to acquire concentration ({\it dhyan}) and to prolong
concentration on an object or a thought ({\it dharna}). 

After this attainment one can reach a stage where one's awareness of self
and the surrounding is at its best.  Then in its last stage, Yoga
prescribes one's acute awareness to be decontextualized\cite{rao} from all
perceptions limited by spatio-temporality and thus to reach a pinnacle
called ({\it samadhi}) where one attains an understanding of everything
and has no doubts. {\em In this sense the Yogic philosophy believes that pure
consciousness transcends all perceptions and awareness}. It is difficult to
understand this on the basis of day to day experience. Why does one need
to sharpen one's awareness to its extreme if one is finally going to
abandon its use ? How does abandonning one's sharpened awareness help in
attaining a realisation that transcends spatio-temporality? Can any one
realise anything that is beyond the space, time and causality ? What is
the purpose of such a consciousness that lies beyond the confinement of
space and time ? 
\end{itemize}
\subsection{Non-Self}

The Non-Self belongs to an external world consisting of others, both
living and non-living.  In the following we discuss whether one can direct
one's consciousness towards others such that one can affect their
behaviour. 
\begin{itemize}
\item{\bf Hypnosis, ESP etc...}

It is a well-known fact that it is possible to hypnotise a person and 
then to make contact with his/her sub-conscious mind. Where does this 
sub-conscious lie ? What is its relation to the conscious mind ? The 
efficacy of the method of hypnosis in curing people of deep-rooted 
psychological problems tells us that we are yet to understand the 
dynamics of the human brain fully.

The field of Para-Psychology deals with `phenomena' like Extra Sensory
Perception (ESP) and telepathy etc. where one can direct one's
consciousness to gain insight into future or to influence others mind. It
is not possible to explain\cite{rao} these on the basis of the known laws
of the world. It has been claimed that under hypnosis a subject could
vividly recollect incidents from the previous lives including near-death
and death experiences which is independent of spatio-temporality. Then, it
is not clear, why most of these experiences are related to past ?  {\em If
these phenomena are truely independent of space and time, then studies
should be made to find out if anybody under hypnosis can predict his/her
own death, an event that can be easily verifiable in due course of time,
unlike the recollections of past-life !} 
\item{\bf PK, FieldREG etc.}

Can mind influence matter belonging to outside of the body ? The studies
dubbed as Psycho-Kinesis (PK) have been conducted to investigate the
`suspect' interaction of the human mind with various material objects such
as cards, dice, simple pendulum etc.  An excellent historical overview of
such studies leading upto the modern era is available as a review paper,
titled `` The Persistent Paradox of Psychic Phenomena: An Engineering
Perspective", by Robert Jahn of Princeton University published in Proc.
IEEE (Feb. 1982). 

The Princeton Engineering Anomalies Research (PEAR) programme of the
Department of Applied Sciences and Engineering, Princeton University, has
recently developed and patented a `Field REG' (Field Random Event
Generator) device which is basically a portable notebook computer with a
built-in truely random number generator (based on a microelectronic device
such as a shot noise resistor or a solid-state diode) and requisite
software for on-line data processing and display, specifically tailored
for conducting `mind-machine interaction' studies. 

After performing large number of systematic experiments over the last two
decades, the PEAR group has reported\cite{srinivasan} the existence of
such a consciousness related mind-machine interaction in the case of
`truely random devices'. They attribute it to a `Consciousness Field
Effect'.  They have also reported that deterministic random number
sequences such as those generated by mathematical algorithm or
pseudo-random generators do not show any consciousness related anomalous
behaviour.  Another curious finding is that `intense emotional resonance'
generates the effect whereas `intense intellectual resonance' does not ! 
{\em It is also not clear what is the strength of the `consciousness 
field' in comparison to all the four known basic force fields of nature}. 
\end{itemize}

One should not reject outright any phenomenon that cannot be explained by 
the known basic laws of nature. Because each such phenomenon holds the 
key to extend the boundary of our knowledge further. But before accepting 
these effects one should filter them through the rigours of scientific 
methodology. In particular, the following questions can be asked:
\begin{itemize}
\item Why are these events rare and not {\em repeatable} ?
\item How does one make sure that these effects are not manifestations of 
yet {\em unknown} facets of the {\em known} forces ?
\item Why is it necessary to have {\em truely random} processes ? How does 
one make sure that these are not merely statistical artifacts ?
\end{itemize}

If the above effects survive the scrutiny of the above questions (or
similar ones) then they will open up the doors to a new world not yet
known to science. In such a case how does one accomodate them within the
existing framework of scientific methods ? If these effects are confirmed
beyond doubt, then one has to explore the possibility that at the
fundamental level of nature, the laws are either different from the known
physical laws or there is a need to complement the known physical laws
with a set of non-physical laws ! In such a situation, these `suspect'
phenomena might provide us with the valuable clue for modifying BE to get
M(BE) that is the basis of everything including both physical and mental ! 
\section{Is there a need for a change of paradigm ?}

Although reductionist approach can provide us with valuable clues
regarding the attributes of consciousness, it is the holistic approach
that can only explain consciousness. But the dualism of
Descarte\cite{narasimhan} that treats physical and mental processes in a
mutually exclusive manner will not suffice for understanding consciousness
unless it makes an appropriate use of complementarity for mental and
physical events which is analogous to the complementarity evident in the
quantum world. 
\section{Conclusion}

Where does the brain end and the mind begin ?  Brain is the physical means
to acquire and to retain the information for the mind to process them to
find a `meaning' or a `structure' which we call `understanding' that is
attributed to consciousness. Whereas attributes of consciousness can be
reduced to BE [or to M(BE)], the holistic process of consciousness can
only be understood in terms of `quantum communication', where
`communication' has an appropriate meaning. Maybe one has to look for 
such a suitable definition of communication at the level of `quantum 
vacuum'.
\section{Acknowledgements}

It is a pleasure to thank the organisers, in particular to Prof. B. V. 
Sreekantan and Dr. Sangeetha Menon; for the hospitality and encouragement
as well as for providing the conducive atmosphere that made this article
possible.

\end{document}